\documentclass{iucr}

\usepackage{siunitx}
\usepackage{color}
\usepackage{mathtools}
\usepackage[normalem]{ulem}
\usepackage{adjustbox}
\usepackage{array}
\usepackage{booktabs}
\usepackage{multirow}
\usepackage{siunitx}

\usepackage{amsmath}
\usepackage{url}

\DeclareMathOperator{\sinc}{sinc}

\definecolor{JSR_blue}{RGB}{51, 102, 154}

\newcolumntype{R}[2]{%
    >{\adjustbox{angle=#1,lap=\width-(#2)}\bgroup}%
    l%
    <{\egroup}%
}

\makeatletter
\@ifclasswith{iucr}{preprint}{

}{

}
\makeatother

 \journalcode{X}              

\begin{document}                  


\title{Ray tracing simulations of bending magnet sources with SHADOW4}

\cauthor[a]{Manuel}{Sanchez del Rio}{srio@esrf.eu}{address if different from \aff}
\author[b]{Howard A.}{Padmore}

\aff[a]{European Synchrotron Radiation Facility, 71 Avenue des Martyrs F-38000 Grenoble, France}
\aff[b]{Advanced Light Source LBNL, Berkeley CA, USA}



\maketitle                        


\begin{synopsis}
XXXX to be done.
\end{synopsis}


\begin{abstract}
We explain the algorithm employed in SHADOW4 to generate bending magnet sources. We provide an overview of the fundamental equations used to calculate the spectral and angular distributions of synchrotron emission from bending magnets. We outline the procedures for ray sampling. We present examples for intensity and polarization of the bending magnet at the old ESRF1 storage ring, and phase spaces for the bending magnet of the upcoming ALSU ring. Finally, we discuss the calculation of the effective source size and compare it with ray tracing.
\end{abstract}

\section{Introduction}
\label{sec:introduction}

Ray tracing simulations serve as essential tools for designing, optimizing, and analyzing optical systems, providing invaluable insights into how light rays propagate through intricate arrangements of optical components. In this paper we describe the algorithm used to sample rays according to the emission of a bending magnet source, and create the source beam for further ray tracing. 

SHADOW included a model for bending magnet (BM) from its first version SHADOW1 \cite{Cerrina1984}. This code was used in SHADOW2 and SHADOW3 \cite{codeSHADOW} without  modifications or upgrades. 

This paper aims to describe the methods and algorithms used in SHADOW4\footnote{Available at: {\tt https://github.com/oasys-kit/shadow4}} for simulating bending-magnet sources. 
The BM model has been upgraded to deal with some relevant effects in 4$^\text{th}$ generation synchrotron sources, with extremely small electron sizes in the bends, therefore the effective horizontal photon source size can be massively larger than the electron beam size. SHADOW4 considers the effect of the radiation cone emitted by the electrons also in the horizontal plane. In previous versions of SHADOW this was ignored as its effect was small compared with higher values of electron emittance in old storage rings.


\section{Basic equations of bending magnet emission}
\label{sec:theory}

This section summarises the basic equations used for calculating the bending magnet emission. They are coded in different functions available in the {\tt srfunc} module of the {\tt sr-xraylib} package\footnote{Available at {\tt https://github.com/oasys-kit/sr-xraylib}}. 

A bending magnet is a dipole creating a magnetic field $B$ that is supposed to be constant over a length $L$. An electron that enters in the bending magnet with a linear velocity $\beta$ is forced to describe a circular trajectory. The radius of curvature is (SI units):
\begin{equation}
    R = \frac{p}{e B} = \frac{m_0 \beta \gamma c}{eB}; \text{or,~} R[m] \approx 3.3356 \frac{E[GeV]}{B[T]}.
\end{equation}
where $c$ is the velocity of the light, $e$ is the electron charge, $B$ is the magnetic field, $m_0$ is the electron mass at rest, $\beta$ is the electron velocity in $c$ units, and $\gamma$ is the electron energy in units of the electron energy at rest:
\begin{equation}\label{eq:gamma}
    \gamma = \frac{E_e}{m_0 c^2} \text{~or;~} \gamma = 1956.95 ~ E_e[GeV] 
\end{equation}

The electron or electron beam performs a circular trajectory limited in length by $L$. Therefore, the change of angle is changed by $\Delta\theta=L/R$. Along this curved trajectory the electron emits radiation losing a total energy (over a complete revolution)
\begin{equation}
    \delta E = \frac{e^2 \beta^3 \gamma^4}{3 \epsilon_0 R}, \text{or, }
    \delta E [keV] = 88.4715 \frac{[E[GeV]]^4}{R[m]}
\end{equation}

The critical energy divides the power spectrum into two parts each one with one-half of the total power. It is
\begin{equation}
    E_c = \frac{3~\hbar \gamma^3 c}{2 R} \text{; or,~} E_c[keV] = 2.2183 \frac{E[Gev]^3}{R[m]}.
\end{equation}

\begin{figure}
    \centering
    \includegraphics[width=0.98\textwidth]{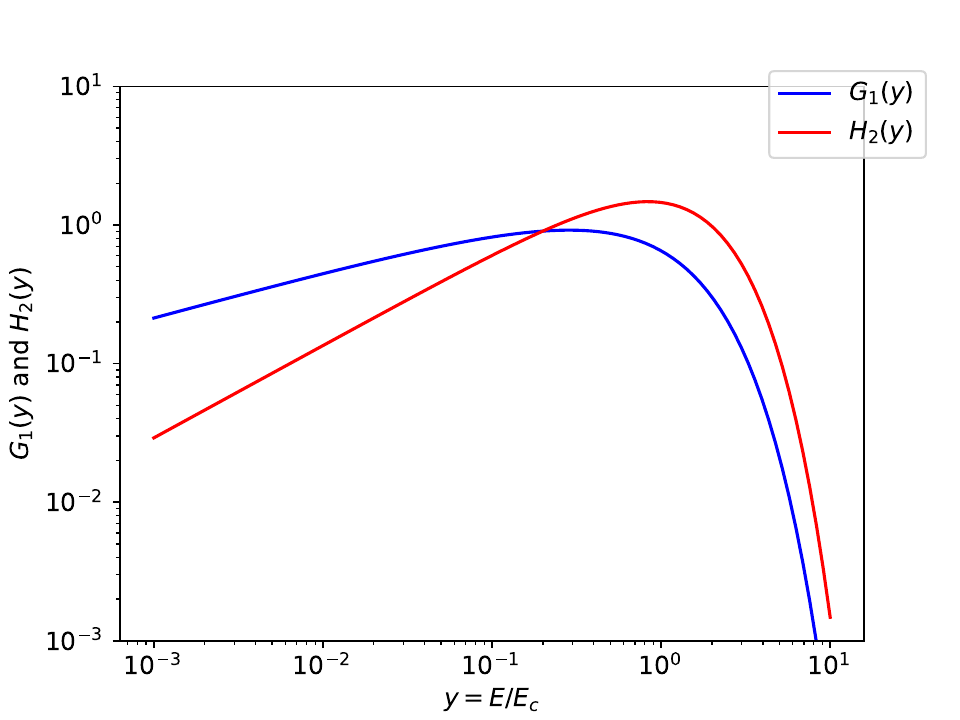}
    \caption{Universal functions for synchrotron emission (equations~\ref{eq:G1},\ref{eq:H2}). The maximum of $G_1$ is attained at $y \approx 1/3$.}\label{fig:universal_functions}
\end{figure}

The emission of the bending magnet radiation are expressed as a function of ``universal" functions (see   Fig.~\ref{fig:universal_functions}),  {\tt sync\_g1} [eq. 20 in \cite{Green1976}] 
\begin{equation}\label{eq:G1}
    G_1(y) = y \int_y^\infty K_{5/3}(x) dx,
\end{equation}
and  {\tt sync\_hi} [eq. 21 in \cite{Green1976}]
\begin{equation}\label{eq:H2}
    H_2(y) = y^2 K^2_{2/3}(y/2).
\end{equation}

The spectrum of the BM is a function of the photon energy and vertical angle. 
The spectrum on-axis (vertical angle $\psi=0$)
is given by [eq. 23 in \cite{Green1976}, or eq. 3 in \cite{xraydatabooklet}] 
\begin{multline}
N_0[\text{photons}/s/0.1\%bw/mrad^2] =  \\
    1.32641~10^{13} (E[GeV])^2 I[A] H_2(E/E_c).   
\end{multline}
The spectrum fully integrated over $\psi$ [eq. 23 in \cite{Green1976}, or eq. 5 in \cite{xraydatabooklet}]
\begin{multline}
    N[\text{photons}/s/0.1\%bw/mrad] = \\
    2.46046~10^{13} E[GeV] I[A] G_1(E/E_c). 
\end{multline}

\begin{figure}
    \centering
    \includegraphics[width=0.85\textwidth]{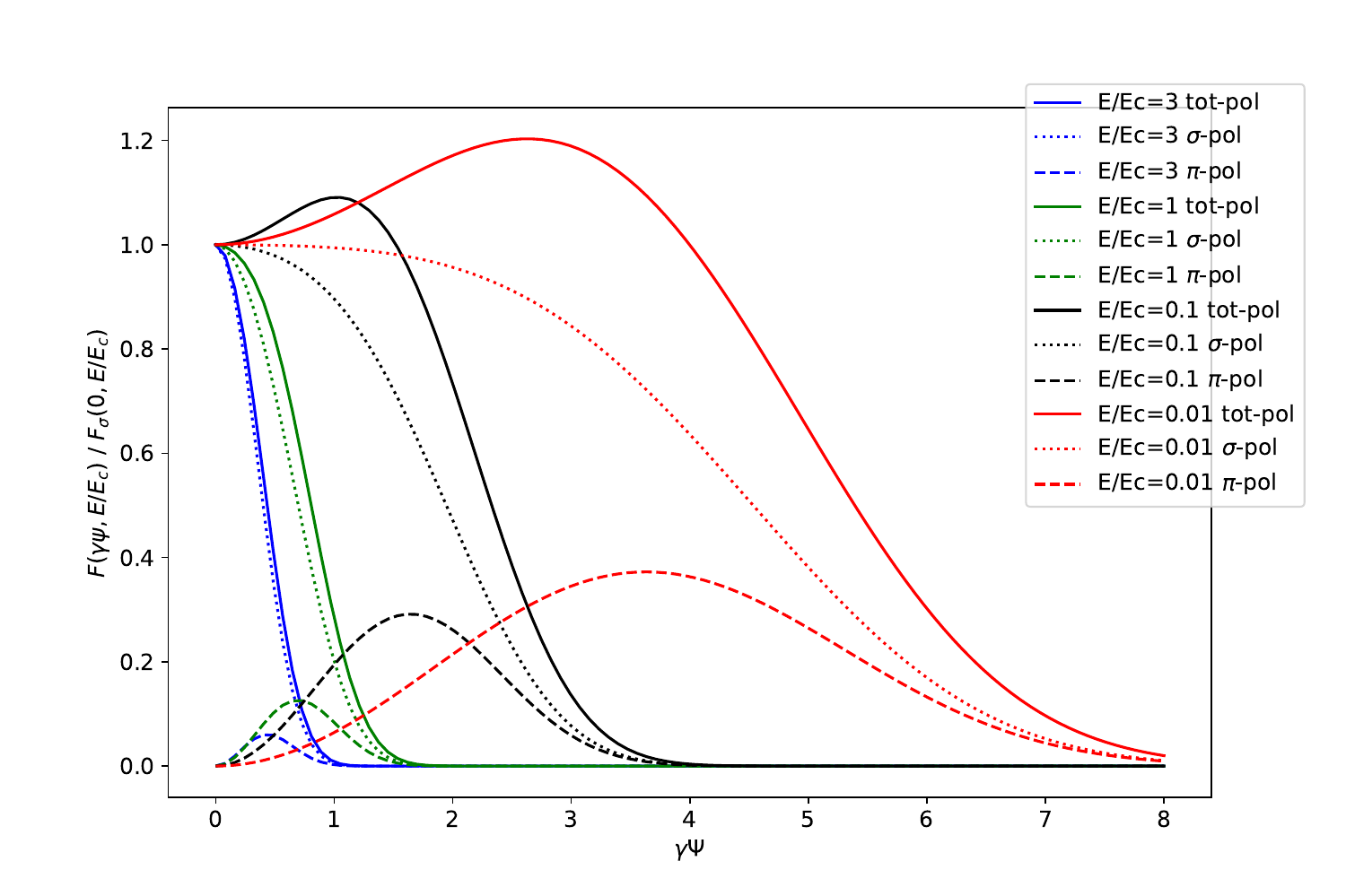}
    \caption{Angular emission for the synchrotron radiation for different photon energies $\epsilon=E/E_c$ at $\sigma$ polarization $F_\sigma(\Psi,\epsilon)/(F_\sigma(0,\epsilon))$ and $\pi$ polarization $F_\pi(\Psi,\epsilon)/(F_\sigma(0,\epsilon)$.
    }\label{fig:angular_emission}
\end{figure}

For each photon energy $E$, the radiation emitted by the electron beam at a point of its trajectory is a cone, with angular distribution following the function [eq. 11 in \cite{Green1976}]
\begin{equation}\label{eq:f}
    F(\Psi,\epsilon) =
    \left[ (1+\Psi^2) \left( l_2 K_{2/3}(\xi) + l_3 K_{1/3}(\xi) \frac{\Psi}{\sqrt{1+\Psi^2}},\right)\right]^2
\end{equation}
where $\Psi=\gamma\psi$ is the reduced angle, $\epsilon=E/E_c$ is the reduced energy and $\xi=(1+\Psi^2)^{3/2} \epsilon / 2$, and $l_{2,3}$ define the polarization state. The total flux is [adding the $\sigma$-polarization ($l_2$ = 1, $l_3$ = 0) plus the $\pi$-polarization components ($l_2$ = 0, $l_3$ = 1)]
\begin{equation}
    N(\psi, E) = N_0(E) \frac{F_\sigma(\Psi,\epsilon) + F_\pi(\Psi,\epsilon)}{F_\sigma(0,\epsilon)}, 
\end{equation}
[see a plot of the $F$ function for $\sigma$ and $\pi$ polarizations in Fig.~(\ref{fig:angular_emission})]

\section{Sampling algorithm}
\label{sec:sampling}

Creating a beam for ray tracing implies sampling rays that follow the bending magnet distribution. We define a reference frame centered in the center of the bending magnet, with $y$ along the electron beam direction, $x$ in horizontal and $z$ in vertical. The $i-th$ ray will be defined by the positions $\textbf{x}_i=(x_i,y_i,z_i)$, direction $\textbf{v}_i$ and energy $E_i$. The ray intensity is defined via the electric vectors for the two polarizations: $\textbf{E}_\sigma$ and $\textbf{E}_\pi$. Shadow defines a ray with intensity one at the source position; $|\textbf{E}_\sigma|^2 + |\textbf{E}_\pi|^2=1 $. The user selects the photon energy limits $E_{min}$ and $E_{max}$. A monochromatic source for a photon energy holds  $E=E_{min}=E_{max}$. 
The sampling algorithm goes through different steps:

\begin{enumerate}
    \item Sample photon energy and SR emission angles $\psi_i$. For a monochromatic source, use the 1D distribution $N(\psi,E)$ [equation~(\ref{eq:f})] to get sampled values of the radiation angle in both the vertical and the horizontal directions ($\psi_{V,i}$ and $\psi_{H,i}$, respectively). For a polychromatic source, use the 2D distribution $N(\psi,E)$ to obtain sampled values of $E_i$, $\psi_{V,i}$, and $\psi_{H,i}$. 
    \item sample the coordinates of the rays along the ideal trajectory. The trajectory is an arc and all its points have the same probability of emission. Therefore we sample the angles $\theta$ in the uniform distribution of width $\Delta \theta$, and compute the coordinates $(R \cos\theta, R \sin\theta, 0)$ and direction $(\theta_i, \sqrt{1-\theta_i^2},0)$.
    \item Sample the angle and coordinates due to the electron emittance. The electron emittance is given by the beam sizes $\sigma_{x,z}$ and beam divergences $\sigma_{x',z'}$ at the waist (where $<x x>_w=\sigma_x^2$, $<x x'>_w$=0, $<x' x'>_w=\sigma_{x'}^2$). In the general case that the waist is not at the center of the BM, its coordinate position (with origin at the center of the BM) $w_{x,z}$  is entered ($w_{x,z} > 0 $ means that the waisrt is found downstream from the BM). The covariance matrix in H for a ray with $y_i$ coordinate 
    \begin{subequations}
        \begin{align}
    &C_x = 
    \begin{bmatrix} 
    <x x> & <x x'>\\
    <x x'> & <x' x'> 
    \end{bmatrix} 
    = \\ 
    &\begin{bmatrix} 
    <x x>_w + (y_i + w_x)^2 <x' x'>_w  & (y_i + w_x) <x'x'>_w\\
    (y_i + w_x) <x'x'>_w & <x' x'>_w 
    \end{bmatrix}
        \end{align}
    \end{subequations}
    A similar expression is obtained for the vertical direction ($z$). Then sampled values are obtained from these Gaussian 2D distributions. Note that there is a correlation in the coordinate and direction space.
    \item The final ray coordinates are obtained by adding the coordinates in the ideal electron trajectory (arc) plus the sampled values for the electron emittance. 
    \item The final ray angles are obtained by adding the angles from the ideal electron trajectory (arc) plus the sampled values for the electron emittance plot the sampled values from the SR emission ($\psi_i$). From this the direction (unitary) vector is constructed.
    \item The polarization degree of each rays is calculates as
    \footnote{Note that SHADOW defines the polarization degree on the amplitude, not on the intensity.}
    \begin{equation}
        P_i = \frac{\sqrt{F_\sigma(\psi_{V,i},E_i)}} {\sqrt{F_\sigma(\psi_{V,i},E_i)}+\sqrt{F_\pi(\psi_{V,i},E_i})}
    \end{equation}
    Using this factor, the electric vectors $\textbf{E}_{\sigma,\pi}$ are constructed taking 
    care of the orthogonality of the $\sigma$ and $\pi$ components.
    
\end{enumerate}

Once all these steps are performed, the rays are packed in the object {\tt S4Beam}. 

\vspace{1cm}

\section{Examples of ray tracing bending magnet sources}\label{sec:MLexamples}

\subsection{Intensity and polarization diistributions for the ESRF1 BM}
\begin{figure}
    \centering
    \includegraphics[width=0.85\textwidth]{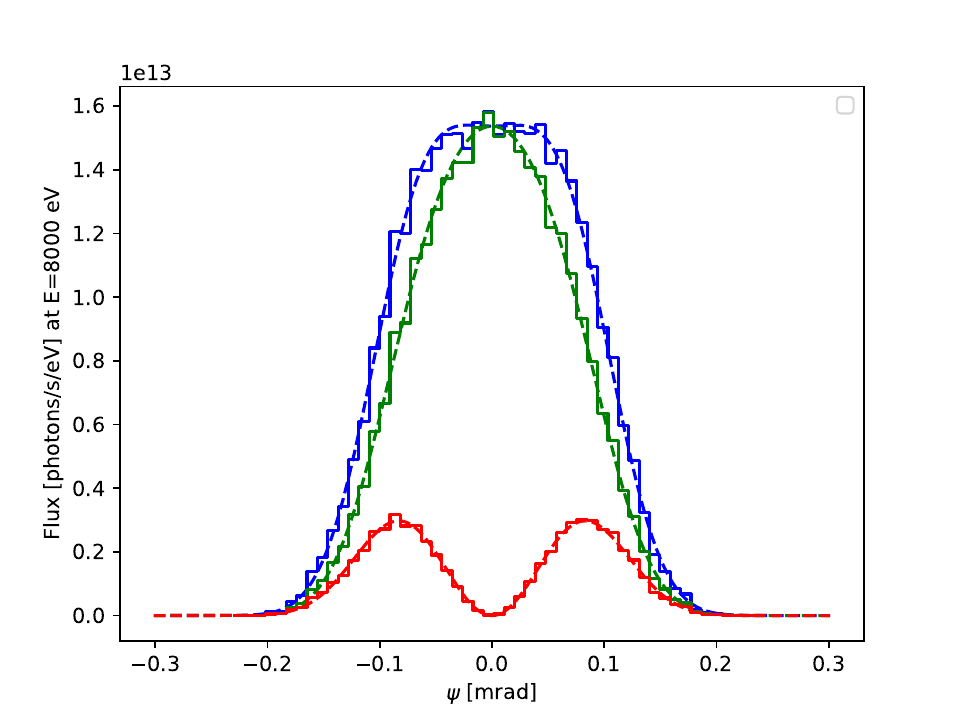}
    \caption{Histograms of the intensity of the rays (total polarization in blue, $\sigma$-polarization in green and $\pi$-polarization in red) versus $\psi$ for the ESRF1 BM source. They are compared with the theoretical flux distribution that has been used in the sampling process. }\label{fig:esrf1_histo_psi}
\end{figure}

We want to illustrate here the emission of the bending magnet. As an example of a hard x-ray BM, we use the parameters of the BM installed at the old ESRF1 storage ring. It had $B$=\SI{-0.8}{T} (the minus sign means that the magnetic field is downwards)
\footnote{In SHADOW the $y$ axis is tangent to the trajectory at the BM center pointing in the direction of the beam, $z$ is vertical pointing upside, therefore the $x$ axis points towards the interior of the ring. With this in mind, the electron velocity $v_y>0$. If $B<0$, $F=-e (\textbf{v} \times B)$ has only $x$ component $F_x>0$, therefore the trajectory points have $x>0$.}
, and $R$=\SI{25.18}{\meter} (unsigned). We use $\Delta\theta$=1mrad divergence, therefore $L=\Delta \theta R$=25.18mm. The electron size and divergences are 
$\sigma_x$=\SI{78}{\micro\meter}, 
$\sigma_z$=\SI{48.7}{\micro\meter}, 
$\sigma_{x'}$=\SI{36}{\micro\radian}, 
$\sigma_{z'}$=\SI{1.06}{\micro\radian}. We first simulated a monochromatic source at 8 keV. In Fig.~\ref{fig:esrf1_histo_psi} we display the histograms of the intensity for the different polarizations (flux) as a function of the vertical direction $z'\approx\psi$. The histograms are calibrated to show the absolute flux. 

\begin{figure}
    \centering
    \includegraphics[width=0.85\textwidth]{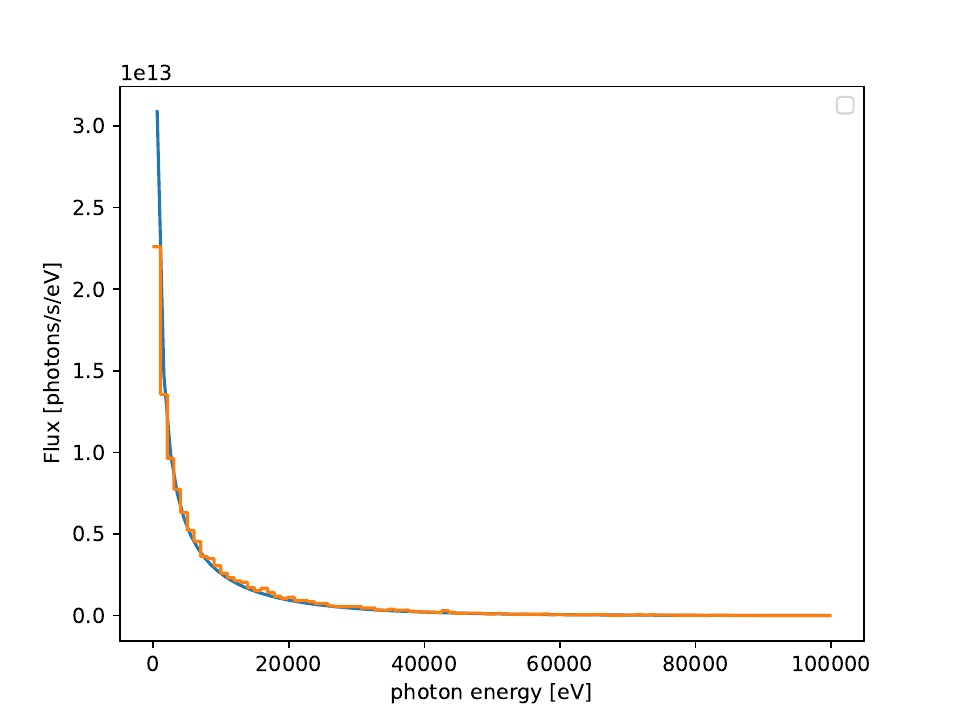}
        \includegraphics[width=0.85\textwidth]{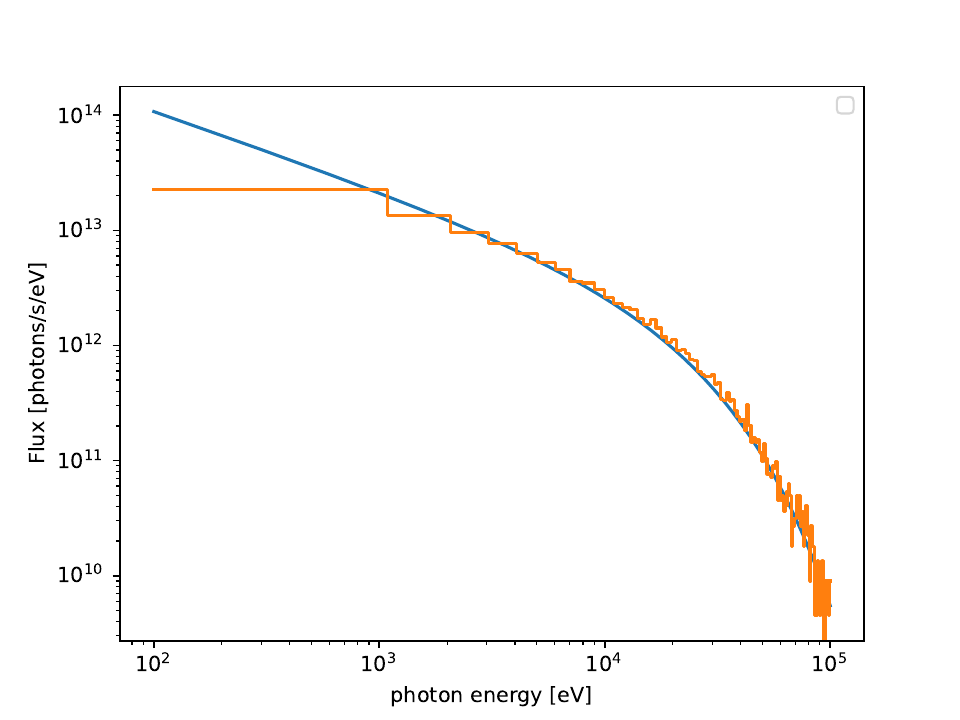}
    \caption{Histograms of the rays intensity versus photon energy for the ESRF1 BM source. They are compared with the theoretical flux distribution. a) linear scales; b) logarithmic scales.
    }\label{fig:esrf1_histo_E}
\end{figure}

A second calculation is done for a polychromatic source, applying the photon energy limits $E_{min}$=100~eV and $E_{min}$=100~keV. The histogram of the flux distribution is presented in Fig.~\ref{fig:esrf1_histo_E}, and compared with the theoretical flux. In SHADOW4 we implemented a mechanism to easily calculate the source spectrum spectrum (with the method {\tt calculate\_spectrum()} of the {\tt S4BendingMagnetLightSource} class) that can be used to normalize the histogram units and display it in absolute flux units.

\subsection{Phase space for a soft x-ray beamline at ALSU}
\begin{figure}
    \centering
    a)~~~~~~~~~~~~~~~~~~~~~~~~~~~~~~~~~~~~~~~~~~~~~~~~~~~~~~~~~~~~~~~~~~~~~~~~~~~~~~~~~~\\
    \includegraphics[width=0.85\textwidth]{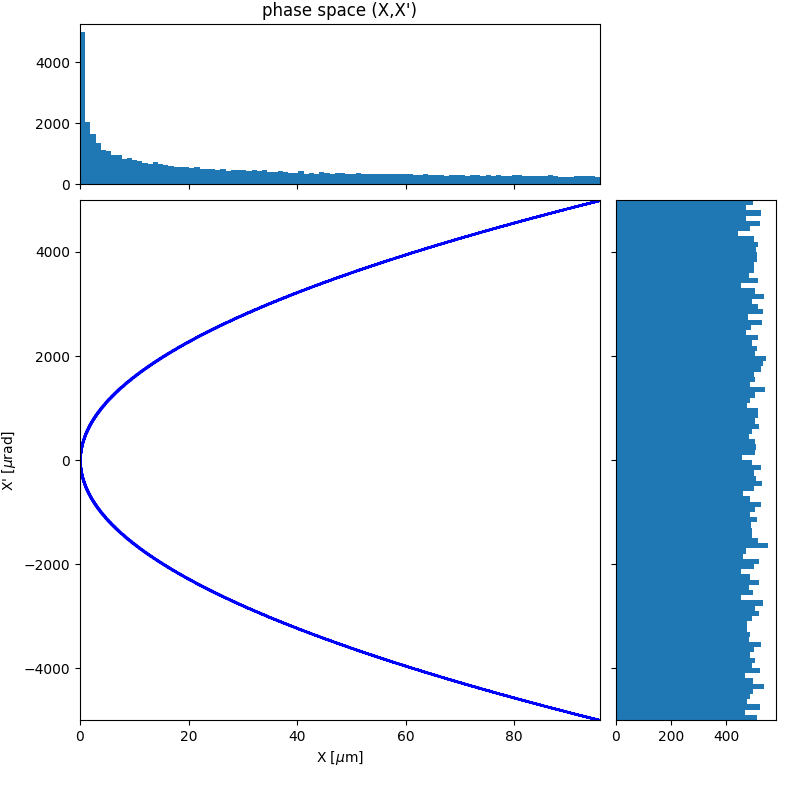}
    b)~~~~~~~~~~~~~~~~~~~~~~~~~~~~~~~~~~~~~~~~~~~~~~~~~~~~~~~~~~~~~~~~~~~~~~~~~~~~~~~~~~\\
    \includegraphics[width=0.85\textwidth]{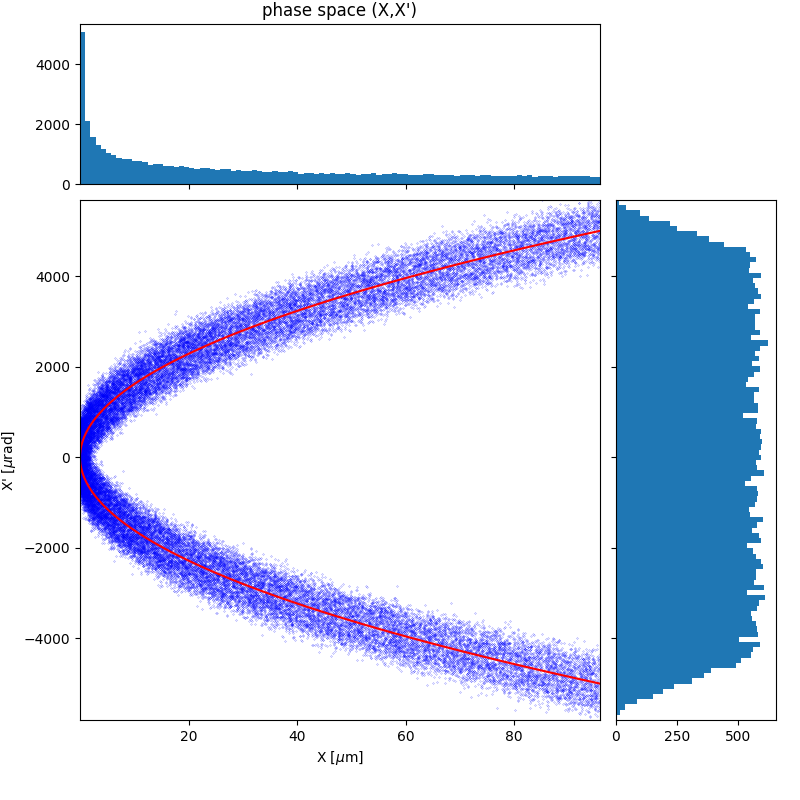}
    \caption{Phase space $(x,x')$ plots for the ALSU bending magnet (no emittance). 
    a) phase space produced of rays emitted tangentially to the electron trajectory.
    b) rays are emitted tangentially to the trajectory with an emission angle that considers the natutal emission of the synchrotron. 
    The plot of the parabola $(R\theta^2/2,\theta)$ is shown in red (see text).
    In the right histogram it can be appreciated the intensity vs $x'$ plot (angular emission).  
    }\label{fig:alsu_phasespace}
\end{figure}

We first illustrate here the horizontal phase space $(x, x')$ in a bending magnet. We used the ALSU case, $E_e$=\SI{2}{\giga\eV}, $B$=\SI{-0.87}{T}, emitting at soft x-ray energies $E$=\SI{282}{\eV} (C K-edge, wavelength \SI{4.4}{\nano\meter}). The horizontal aperture\footnote{
A soft x-ray beamline extending to \SI{1.5}{\kilo\eV} might have a mirror grazing angle of \SI{1.5}{\degree}, and so in the front end, \SI{5.5}{\meter} from the source, for a tangentially focusing horizontally deflecting mirror, the length for \SI{5}{\milli\radian} would be \SI{1.05}{m}. This is practical. 
Higher aperture would be possible with a sagittal focusing mirror deflecting in the vertical (for example paraboloid - plane grating - paraboloid). It may be possible in this way to accept up to \SI{10}{\milli\radian}, and so for an extreme case, we have adopted this value. 
} is $\Delta \theta$=10~mrad.
In the ideal case that there is no electron emittance, the photons will be emitted tangentially along an arc of radius $R$ and length $L \approx R~\Delta\theta$. A point with angle $\theta$ has coordinates $(x,y,z)=(R(1-\cos\theta), R\sin\theta, 0)\approx(R\theta^2/2, R\theta, 0)$, and horizontal deviation or angle $x'\approx \theta$. The horizontal phase space is just a parabola  $(x,x')=(R\theta^2/2, \theta)$ (see Fig~\ref{fig:alsu_phasespace}a).
When considering the natural emission cone of the synchrotron radiation, there is an additional angle to be added to $x'$ (see Fig~\ref{fig:alsu_phasespace}b). Observe that the range of $x$ is the same in Figs~\ref{fig:alsu_phasespace}a and b, because the cartesian coordinates $(x,y,0)$ of the emitting rays are the same. 
Observe that the photon distribution extends to +100 microns, while the electron beam sigma for ALSU is 7 microns. This justifies the fact that we neglected electron emittance for this particular example (note that the electron divergence can be considered in SHADOW4).

\begin{figure}
    \centering
    a)~~~~~~~~~~~~~~~~~~~~~~~~~~~~~~~~~~~~~~~~~~~~~~~~~~~\\
    \includegraphics[width=0.85\textwidth]{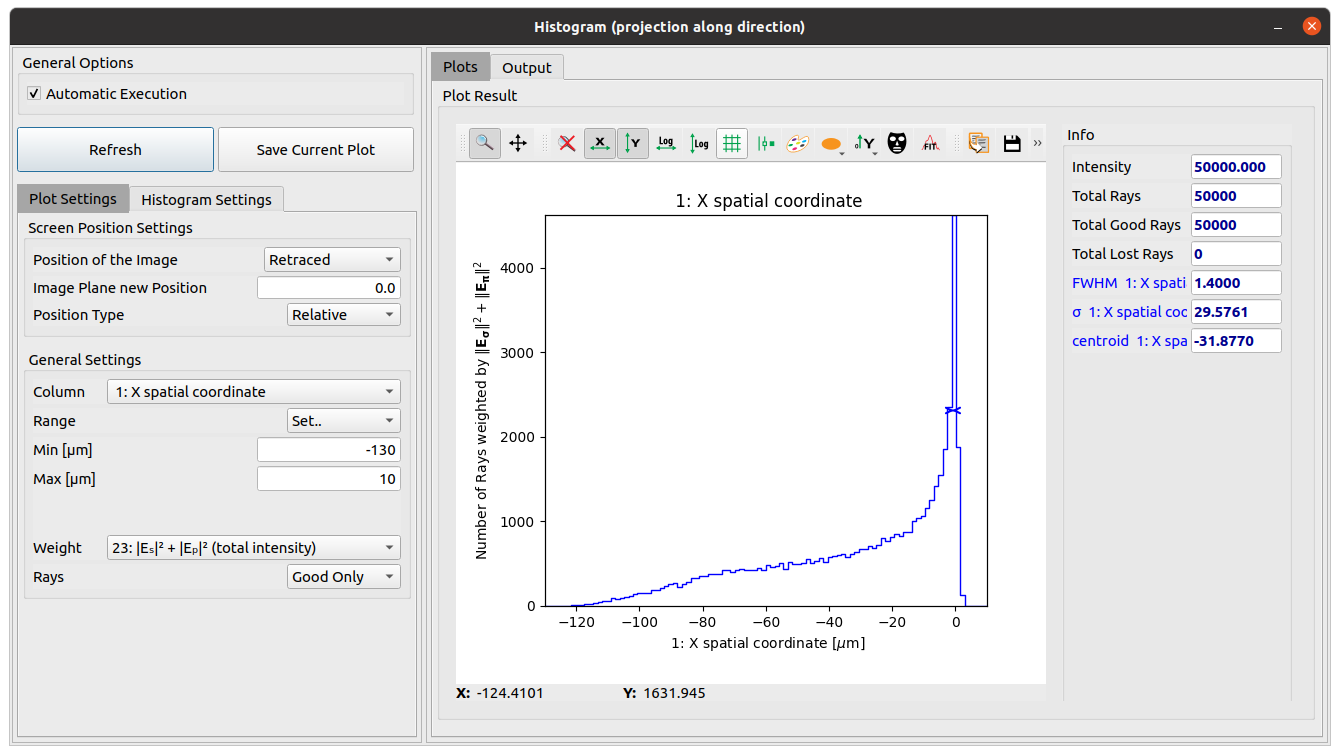}
    b)~~~~~~~~~~~~~~~~~~~~~~~~~~~~~~~~~~~~~~~~~~~~~~~~~~~\\
    \includegraphics[width=0.85\textwidth]{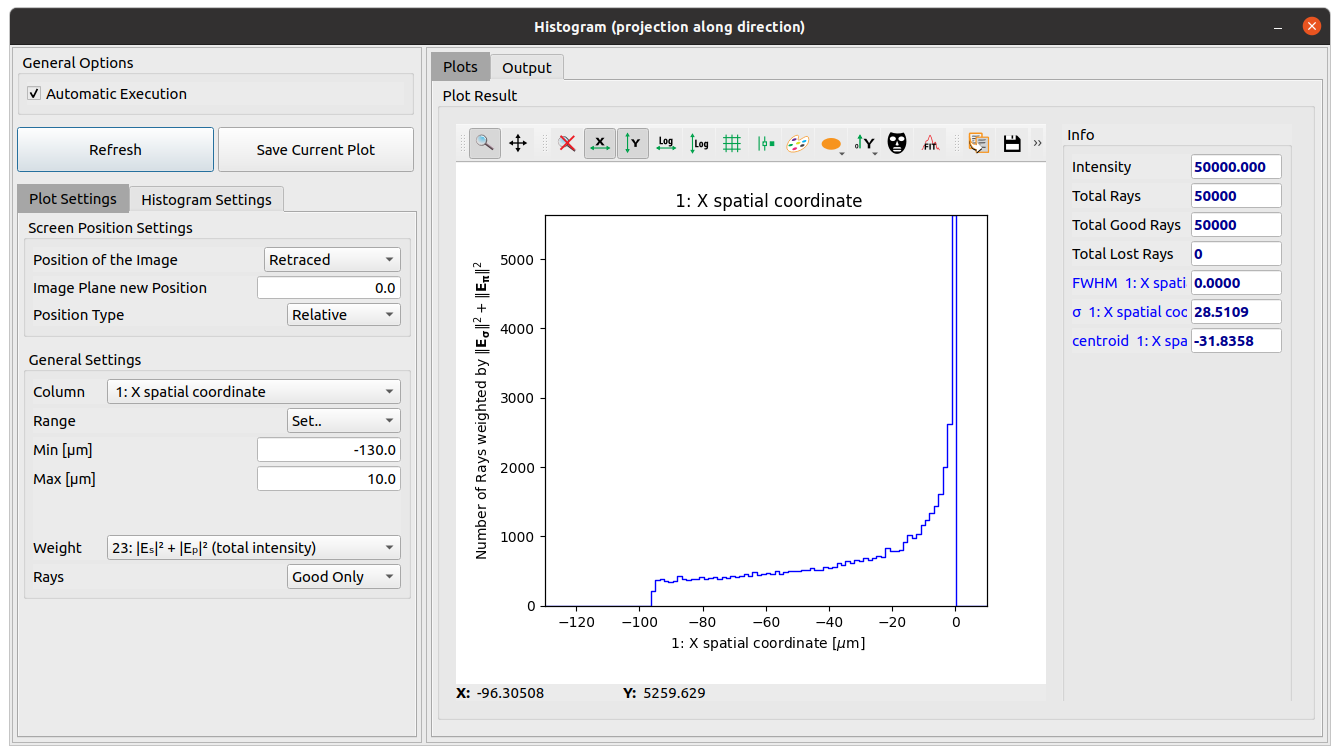}
    \caption{Histogram of the intensity vs $\Delta x$. a) SHADOW4 result, b) old SHADOW3 result that does not include the sampling of the angle of the natural SR emission in the horizontal direction.  
    }\label{fig:alsu_histogram}
\end{figure}

If a photon exits tangentially at that points of the trajectory. it will cross the $x$ axis at a point with coordinate $\Delta x= x' R \theta= R \theta^2$. To display a plot of the intensity vs this projected coordinate $\Delta x$ in SHADOW, we perform a {\tt retrace} of the source beam over a zero distance. The {\tt retrace} method of the {\tt S4Beam} class traces all the rays along their directions to a plane sitting at $y=0$. Then a histogram is performed to compute the intensity (see Fig~\ref{fig:alsu_histogram}).

The {\tt retrace} mechanism is important to define ``effective" sources. In real cases you can image the source with an ideal focusing device (real lens or mirror) in one-to-one magnification. In ray tracing, you can do that, or better, forward-propagate the source to a given distance and back-propagate the beam the same distance. This will create an image of the 3D source at the $y=0$ plane (without depth). The effect of forward-propagate to a plane at a distance $d$ and then back-propagate a distance $-d$ is equivalent to directly propagate to the plane at $y=d-d=0$, as we did in the last paragraph.
Moreover, this is important from the optics points of view.  The plane at $y$=0 is the conjugate plane for a downstream focusing plane (e.g. focusing to a plane where slits are placed).  With magnification of the optic used, one can back transform the slit or aperture back to the source and impose a cut.

\begin{figure}
    \centering
    \includegraphics[width=0.85\textwidth]{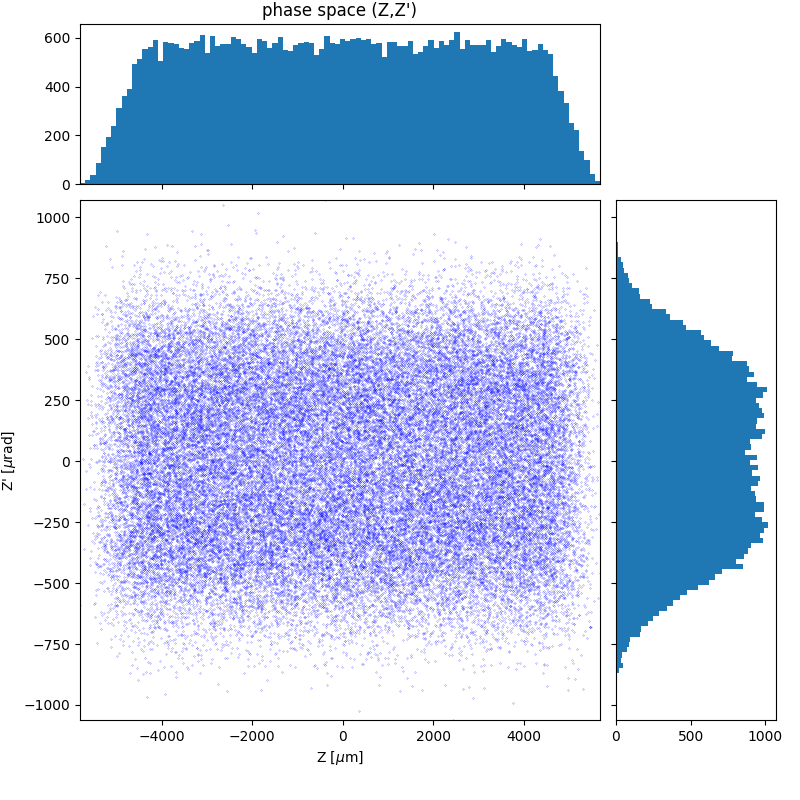}
    \caption{Phase space $(z,z')$ plots for the ALSU bending magnet (no emittance) at a photon energy of \SI{282}{\eV}.
    }\label{fig:alsu_phasespaceV}
\end{figure}

For completeness, Fig~\ref{fig:alsu_phasespaceV} shows the phase space in the vertical direction $(z,z')$.

\subsection{The BM effective source size}
\begin{figure}
    \centering
    \includegraphics[width=0.85\textwidth]{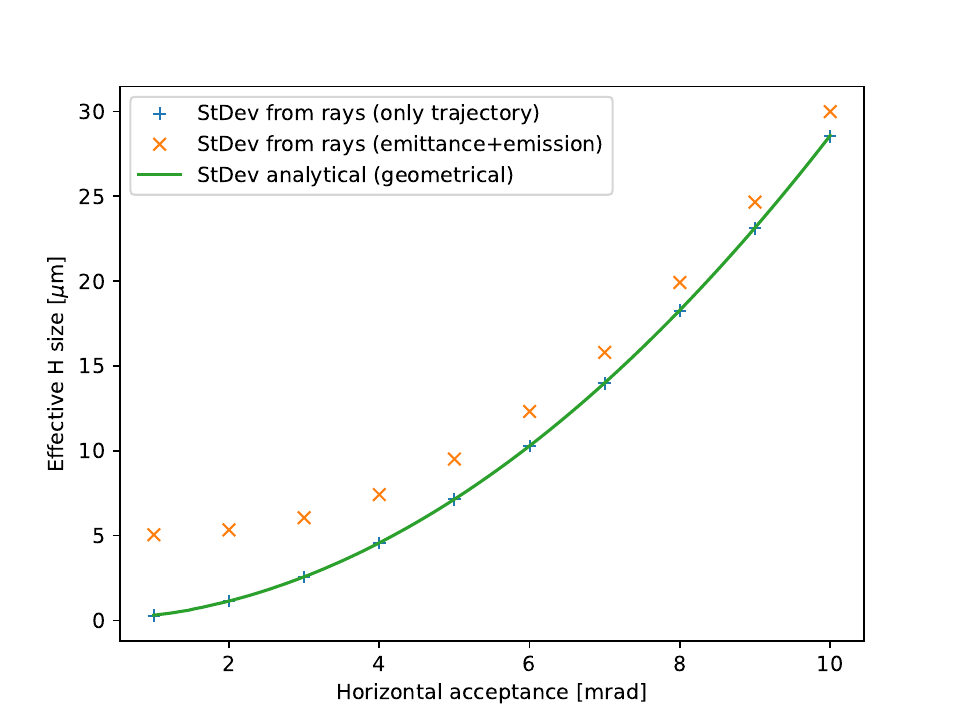}
    \caption{Effective source size (standard deviation) in horizontal due to the geometric effect (trajectory). Calculations from ray tracing (with no emittance and no cone emission) are compared with analytical results from equation~(\ref{eq:nucara}). 
    }
    \label{fig:alsu_std}.
\end{figure}

The source simulated for ray tracing contains the 3D coordinates of the rays (photons). The source has a finite depth (in $y$) due to the electron trajectory over an arc. As introduced before, the ``effective source" is obtained by propagating all rays to the plane $y=0$, making in this way a new ``effective" source with no depth. The effective source size depends on i) the geometry of the electron trajectory from where the photons are emitted, ii) the electron beam size and divergence, and iii) the natural emission cone of the radiation at each point of the trajectory.
In very low emittance storage rings, where the beam size is extremely small, the geometric part can become dominant (as shown in Fig.~\ref{fig:alsu_phasespace}). 
The effective source distribution, as shown in Fig~\ref{fig:alsu_histogram} has a particular shape, peaked at the origin and a long tail on one size. This shape depends on the horizontal angular acceptance, and the width can be estimated by the standard deviation of the ray's coordinates $x$ for the beam projected on the $y=0$ plane. For the ALSU system under discussion, the standard deviation of the projected $x$ is in Fig.~\ref{fig:alsu_std}. It is compared with the analytical expression \cite{Nucara1995}
\begin{equation}\label{eq:nucara}
    \sigma^{\text{geom}} = R \sqrt{
    \frac{3}{2}+\frac{1}{2} \sinc(\Delta\theta) -
    2\sinc(\Delta\theta/2) -
    \left(1-\sinc(\Delta\theta/2) \right)^2 },
\end{equation}
where $\sinc(x)=(\sin x)/x$.

Making ray tracing simulations with emittance (electron size $\sigma_{x,z}$=\SI{7}{\micro\meter}, $\sigma_{x',z'}$=$\epsilon_{x,z}/\sigma_{x,z}$ with emittance $\epsilon_{x,z}$=\SI{70}{\pico\meter\radian}) and radiation emission we show that the effective source size increases (markers ``x" in Fig.~\ref{fig:alsu_std}). 
If we suppose that all contributions are Gaussian, the effectice source size in horizontal is obtained by convolution of the individual Gaussians:
\begin{equation}\label{eq:convolution}
    \Sigma_x^2 = \sigma_x^2 + \sigma_r^2 + \sigma^{\text{geo},2}_x,
\end{equation}
where $\sigma_x$ is a component due to the electrons, $\sigma_r$ is due to the radiation cone emitted from each point along the electron trajectory, $\sigma^{\text{geo}}_x$ is due to the depth (arc trajectory), and $\sigma^{\text{DL}}_x$ is a size contribution due diffraction limit (this is not modeled in SHADOW). 
A similar equation is obtained the vertical direction $z$ with $\sigma^{\text{geo}}_z$=0.

\begin{figure}
    \centering
    \includegraphics[width=0.85\textwidth]{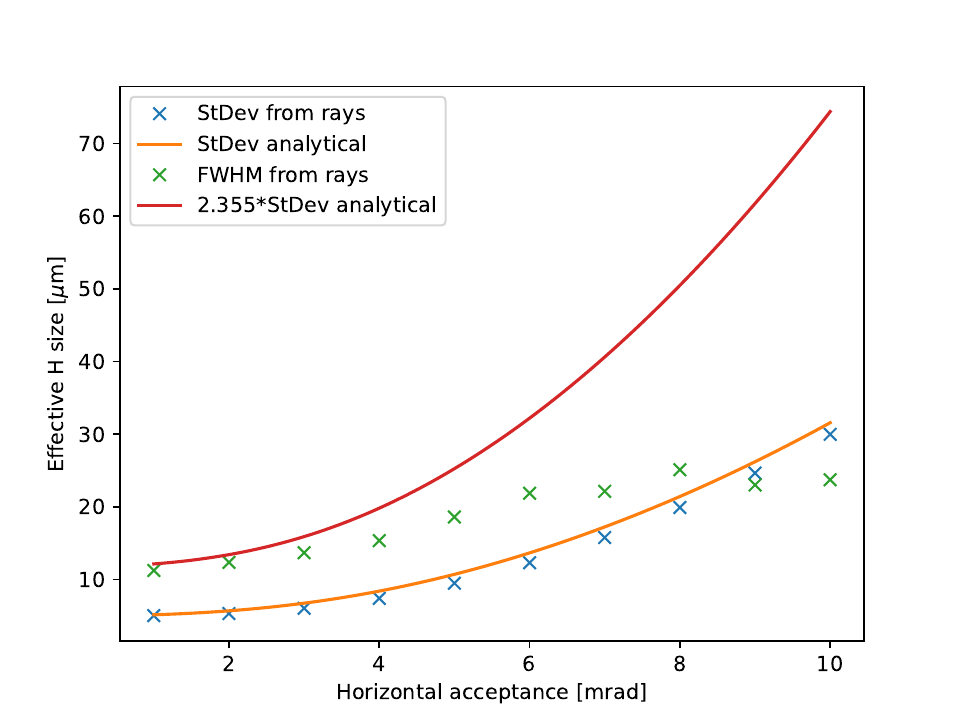}
    \caption{Effective source size (standard deviation and FWHM) in horizontal. Calculations from ray tracing are compared with analytical results from equation~(\ref{eq:convolution}). It shows good agreement for the standard deviation, but lacks agreement for the FWHM.
    }\label{fig:convolution}
\end{figure}

In Fig.~\ref{fig:convolution} we display the comparison of horizontal size (in standard deviation and FWHM) obtained by ray tracing with the results of equation~(\ref{eq:convolution}) (where $\sigma_r$=\SI{163}{\micro\meter}). Both agree for the standard deviation values, but the FWHM(Full Width at Half-Maximum) are very different for high values of $\Delta \theta$, where the effect of the trajectory appears with its highly asymmetric profile, thus not well approximated by a Gaussian. 
 
Last, but not least, we note that in case of radiation of very low energy (e.g. infrared) in small emittance rings, there is another contribution in equation~(\ref{eq:convolution}) due to the diffraction limit size.
It can be estimated as $\sigma^{\text{DL}}=\lambda/(4 \pi \sigma_r)$ with $\lambda$ the photon wavelength and $\sigma_r$ the natural emission angle.
For our case $\sigma_r$=\SI{440}{\micro\radian} and $\lambda$=\SI{4.4}{\nano\meter}, we get $\sigma^{\text{DL}} \approx$\SI{0.8}{\micro\meter}. This effect is not included in SHADOW for the BM source (it is included for the undulator). 

\section{Summary and conclusions}
\label{sec:summary}
 
We summarize the theoretical results of the BM emission and the implementation in SHADOW4 to simulate bending magnet sources. We illustrate its use with three examples, discussing the flux (angular and spectral dependence), the phase space, and the effective source size.


\appendix

\bibliography{iucr}
\bibliographystyle{iucr}


\end{document}